\documentclass[journal]{IEEEtran}
\usepackage{cite}
\usepackage{graphicx}
\usepackage{epstopdf}
\usepackage{amsmath,amssymb,amsfonts}
\usepackage{changes}
\usepackage{textcomp}
\usepackage{xcolor}
\usepackage[ruled,lined,linesnumbered]{algorithm2e}
\usepackage{tabularx,booktabs}
\usepackage[flushleft]{threeparttable}

\usepackage{ragged2e}
\usepackage{multirow}
\usepackage{subcaption}
\usepackage{mathtools}
\usepackage{bm}
\usepackage[colorlinks,
            linkcolor=black,
            anchorcolor=black,
            citecolor=black
            ]{hyperref}

\def\BibTeX{{\rm B\kern-.05em{\sc i\kern-.025em b}\kern-.08em
    T\kern-.1667em\lower.7ex\hbox{E}\kern-.125emX}}

\begin{document}

\title{Cache-enabled Generative Joint Source-Channel Coding for Evolving Semantic Communications}

\author{
Shunpu Tang,
Qianqian Yang,
Jihong Park,
Zhaoyang Zhang,
Kaibin Huang,
and Deniz Gündüz
\thanks{This work was in part presented at the IEEE International Symposium on Personal, Indoor and Mobile Radio Communications (PIMRC), Valencia, Spain, 2024 \cite{Shunpu_PIMRC}.}
\thanks{This work is partly supported by the National Key R\&D Program of China under Grant No. 2024YFE0200802, partly by NSFC under grant No.62293481, No.62201505 and No.625B2165, partly supported by the China Scholarship Council (No. 202406320381).}
 \thanks{ S. Tang, Q. Yang and Z. Zhang are all with the College of Information Science and Electronic Engineering, Zhejiang University, Hangzhou, China (e-mails: \{tangshunpu, qianqianyang20, ning\_ming\}@zju.edu.cn).}
 \thanks{J. Park is with the Information Systems Technology and Design (ISTD) Pillar, Singapore University of Technology and Design (SUTD), Singapore. (e-mail: jihong\_park@sutd.edu.sg)}
 \thanks{K. Huang is with the Department of Electrical and Electronic Engineering,
The University of Hong Kong, Hong Kong (e-mail: huangkb@hku.hk).}
 \thanks{ D. Gündüz is with the Department of Electrical and Electronic Engineering, Imperial College London, London, UK, SW7 2AZ. (Email: d.gunduz@imperial.ac.uk).}
 }

\maketitle
\thispagestyle{empty}
\pagestyle{empty}
\begin{abstract}

Learning-based semantic communication (SemCom) has recently emerged as a promising paradigm for improving the transmission efficiency of wireless networks. However, existing methods typically rely on extensive end-to-end training, which is both inflexible and computationally expensive in dynamic wireless environments. Moreover, they fail to exploit redundancy across multiple transmissions of semantically similar content, limiting overall efficiency. To overcome these limitations, we propose a channel-aware generative adversarial network (GAN) inversion-based joint source-channel coding (CAGI-JSCC) framework that enables training-free SemCom by leveraging a pre-trained SemanticStyleGAN model. By explicitly incorporating wireless channel characteristics into the GAN inversion process, CAGI-JSCC adapts to varying channel conditions without additional training. Furthermore, we introduce a cache-enabled dynamic codebook (CDC) that caches disentangled semantic components at both the transmitter and receiver, allowing the system to reuse previously transmitted content. \textcolor{black}{This semantic-level caching can continuously reduce redundant transmissions as experience accumulates.} Extensive experiments on image transmission demonstrate the effectiveness of the proposed framework. In particular, our system achieves comparable perceptual quality with an average bandwidth compression ratio (BCR) of 1/224, and as low as 1/1024 for a single image, significantly outperforming baselines with a BCR of 1/128.

\end{abstract}

\begin{IEEEkeywords}
Semantic communication, wireless cache, GAN, joint source-channel coding.
\end{IEEEkeywords}

\section{Introduction} 
\subsection{Background}
\textcolor{black}{Recently, many emerging applications from extended reality and fully autonomous driving\cite{Autonomous_6G} to digital twins\cite{Digital_Twin1} and holographic communications \cite{Holographic1}, pose new challenges to existing communication systems and infrastructures. These applications require the transmission of high-information-content sources with very low latency and high fidelity.}
To tackle this challenge, semantic communication (SemCom) has emerged as a promising approach to enhance communication efficiency in the forthcoming 6G networks. Specifically, SemCom prioritizes the transmission of semantically valuable information relevant to the task at the receiver side while discarding irrelevant or less critical details\cite{Semantic1,Semantic2,Semantic3}. \textcolor{black}{As a result, SemCom can be considered as a data compression problem with the distortion measure specified by the downstream task. On the other hand, when the communication takes place over a noisy channel, 
we have a joint source-channel coding (JSCC) problem with semantic distortion measures. Although JSCC has been studied extensively \cite{JSCC_Survey}, its performance under realistic source and channel distributions remained extremely limited until deep learning (DL)-based code designs were introduced in recent years. }

\subsection{Related Works and Motivations}
The pioneering DL-based SemCom system for image transmission, namely DeepJSCC \cite{DeepJSCC}, employed neural networks to implement joint source–channel encoder and decoder pairs, which are trained in an end-to-end fashion to minimize the distortion between the original and reconstructed images. In a related study \cite{zhang2022wireless}, the authors proposed transmitting high-level semantic information, such as text descriptions, to enhance image transmission performance of DeepJSCC, particularly in bandwidth-limited scenarios. DeepJSCC was further extended to digital communication systems in \cite{DeepJSCC-Q,bo2024joint}, enhancing its compatibility with existing communication infrastructure. A CSI-feedback mechanism was introduced in \cite{DBLP:journals/jsait/KurkaG20,10559783}  to adapt the system to varying channel conditions. To improve robustness, the authors in \cite{CL_SemCom} introduced a contrastive learning approach, treating corruption during transmission as data augmentation, thereby enhancing performance under low signal-to-noise ratios (SNRs). Other works \cite{yangDeepJointSourceChannel2022, xuWirelessImageTransmission2022} incorporated attention mechanisms to further mitigate the impact of channel variability. For text and speech transmission, the authors in \cite{zhijin_text,zhijin_speech} proposed the first DL-based SemCom system, which outperformed traditional separate source-channel coding schemes. Furthermore, \cite{semantic_MEC} introduced a deep reinforcement learning-based SemCom system for text transmission, which can dynamically adjust semantic encoding decisions based on real-time network status.

To further improve the efficiency of SemCom systems, a promising approach is to harness the capabilities of recently emerged deep generative models\cite{liang2023generative}, which can produce high-dimensional, realistic multimedia content from low-dimensional data. In \cite{GAN_JSCC}, the authors proposed a GAN-based JSCC approach for image transmission, where a pre-trained GAN model is employed at the receiver to reconstruct high-quality images. Specifically, the output of the DeepJSCC decoder is fed into the pre-trained GAN, which significantly enhances the perceptual quality of reconstructed images.
In \cite{han2022semantic}, the authors utilized a deep speech generation model to reconstruct speech from low-dimensional semantic information, drastically reducing the required bandwidth to merely 0.2\% of that achieved by existing works. Another approach, presented in \cite{tianxiao_GAN}, employs the GAN inversion method \cite{Gan_inversion} to acquire a low-dimensional latent representation of images. These representations are then reconstructed at the receiver using a generative function, achieving perceptual quality comparable to existing methods while transmitting much less data. More recent works \cite{yilmaz2023high,tang2024retrieval, chen2023commin, tang2025enabling, 11251350} adopt diffusion models, enabling realistic reconstructions of the source images at the receiver. Collectively, these approaches have significantly enhanced communication efficiency while achieving superior perceptual quality in reconstructed contents.

However, in these existing SemCom approaches, the DL-based transceivers are typically trained offline and then deployed with fixed parameters, even in a dynamic environment. Such a paradigm lacks flexibility, as adapting to new channel conditions requires additional retraining. Moreover, the redundancy across multiple transmissions of similar content has not been fully exploited, preventing communication efficiency from improving through accumulated transmission experience. \textcolor{black}{Therefore, we aim to design a training-free SemCom framework that can adapt to dynamic channels without retraining and can continuously enhance efficiency through past experience.}



\textcolor{black}{
    Motivated by the observation that powerful generative models trained on large-scale datasets can be directly applied to diverse downstream tasks without additional training or fine-tuning\cite{gruver2023large, Zanella_2024_CVPR, Diffusion_SemSeg}, we aim to explore their potential in SemCom as semantic encoders and decoders. Meanwhile, wireless caching technologies have been shown to significantly reduce communication overhead by storing previously transmitted data\cite{cache_1, SHI202192, cache_2,DBLP:journals/tvt/ZhouXZFLNK23,DBLP:journals/tvt/YinOGYWS23}. Inspired by this, we propose deploying cache memories at both the transmitter and receiver to enable semantic-level caching. \textcolor{black}{Since identical raw data (e.g., entire images) rarely recurs across multiple transmissions, caching raw data offers limited benefits. In contrast, disentangled semantic components, such as “eyes,” “eyebrows,” or “mouth shapes” in facial images, show high similarity across transmissions. By caching and reusing these recurring elements, the system can continuously enhance communication efficiency as experience accumulates.}
}

\subsection{Contributions}

\textcolor{black}{In this paper, we introduce a novel training-free SemCom framework that leverages a pre-trained generative model shared between the transmitter and receiver to serve as the semantic encoder and decoder, thereby eliminating the need for end-to-end training. To further reduce transmission overhead, we incorporate semantic-level caching at both ends of the system. To implement this framework, we propose a novel generative joint source-channel coding (JSCC) approach called \textit{channel-aware GAN inversion-based JSCC} (CAGI-JSCC), which exploits a well-established GAN model, SemanticStyleGAN \cite{SemanticStyleGAN}, to extract disentangled semantic representations, \textcolor{black}{in which different dimensions of semantic information correspond to distinct semantic components.}
In addition, the inversion process explicitly incorporates wireless channel characteristics, allowing the system to adapt its semantic representation to varying channel conditions. At the receiver, transmitted samples are directly reconstructed by feeding the received signals into the same pre-trained SemanticStyleGAN.}

Furthermore, we introduce a semantic-level caching mechanism by leveraging the disentangled semantic information obtained from SemanticStyleGAN. \textcolor{black}{We propose a cache-enabled dynamic codebook (CDC) that encodes repeatedly transmitted semantic components to compact indices, thereby reducing the transmission overhead. Specifically, the CDC is implemented by deploying cache memories at both the transmitter and receiver to store previously transmitted semantic components. } When a new sample is to be transmitted, the transmitter first searches its local cache for similar semantic components. If such components exist, the transmitter replaces them with the corresponding indices of these components in the cache. The receiver can then reconstruct the complete semantic representation by retrieving the corresponding components from its local cache using the received indices.  Efficient cache update and replacement strategies are further developed to ensure synchronization and long-term effectiveness of the codebook.
 The main contributions of this paper are summarized as follows:
\begin{itemize}
    \item We propose a novel CAGI-JSCC approach based on the pre-trained SemanticStyleGAN for training-free SemCom. By incorporating wireless channel characteristics into the GAN inversion process, CAGI-JSCC produces channel-correlated latent codes that can be transmitted directly over noisy channels without explicit channel coding, ensuring robustness and adaptability.
    \item We propose a cache-enabled dynamic codebook (CDC) that stores and reuses fine-grained, disentangled semantic components at both the transmitter and receiver. By transmitting compact indices instead of redundant components, CDC reduces overhead, and we introduce SNR-aware update and eviction strategies to maintain synchronized and effective caches.
    \item \textcolor{black}{We conduct extensive experiments on a real image dataset under various channel conditions} to evaluate the proposed framework. Results show that CAGI-JSCC with CDC achieves superior performance in both reconstruction quality and communication efficiency compared to existing SemCom approaches. In particular, our framework achieves comparable perceptual quality with an average bandwidth compression ratio (BCR) of 1/224, and as low as 1/1024 for a single image, significantly outperforming baselines with a BCR of 1/128.
    \end{itemize}

\subsection{Organization}
The remainder of this paper is organized as follows. Section II introduces the preliminaries and background of DeepJSCC for SemCom systems, as well as the SemanticStyleGAN model used in this work. Section III presents the proposed CAGI-JSCC framework. Section IV details the design of the proposed CDC and its integration with CAGI-JSCC. Section V provides extensive experimental results. Finally, Section VI concludes the paper.

\section{Preliminaries}
\subsection{DeepJSCC for SemCom Systems}
We consider the problem of transmitting images over a noisy channel. At the transmitter, let the original image be denoted by $\bm{x} \in \mathbb{R}^{3 \times N_\text{H} \times N_\text{W}}$, where $3$ comes from the number of image channels, and $N_\text{H}$ and $N_\text{W}$ are the height and width of the image, respectively. The source bandwidth of the image is therefore defined as $N=3 \times N_\text{H} \times N_\text{W}$. 

To improve communication efficiency, the transmitter uses a parameterized JSCC encoder $E(\cdot; \bm{\theta})$ to extract the semantic information of the image and generate a real-valued vector of length $2k$, expressed as
\begin{equation}
\tilde{\bm{z}} = E(\bm{x}; \bm{\theta}) = \left[\tilde{z}_1, \tilde{z}_2, \ldots, \tilde{z}_{2k} \right]^\top \in \mathbb{R}^{2k},
\end{equation}
where $\bm{\theta}$ denotes the learnable parameters of the encoder. Vector $\tilde{\bm{z}}$ is then mapped into a $k$-dimensional complex-valued signal by grouping adjacent elements as real and imaginary parts using a transformation function $\mathcal{C}: \mathbb{R}^{2k} \rightarrow \mathbb{C}^{k}$: 
\begin{equation}
\tilde{\bm{z}}_c =\mathcal{C}(\tilde{\bm{z}}) =\left[\tilde{z}_1 + j \tilde{z}_2,\ \tilde{z}_3 + j \tilde{z}_4,\ \ldots,\ \tilde{z}_{2k-1} + j \tilde{z}_{2k} \right]^\top,
\end{equation}
where $\tilde{\bm{z}}_c \in \mathbb{C}^k$ is the resulting signal. The corresponding inverse mapping $\mathcal{C}^{-1}: \mathbb{C}^{k} \rightarrow \mathbb{R}^{2k}$ converts a complex-valued vector back to the real-valued latent space by stacking its real and imaginary parts. \textcolor{black}{Following the approach in \cite{DeepJSCC}, we introduce a power normalization layer to enforce the transmit power constraint $\bar{P}$, given by
\begin{equation}
    \label{eq::PN}
\bm{z}=\mathrm{PN}(\tilde{\bm{z}}_c)
= \tilde{\bm{z}}_c
\sqrt{\frac{k\bar{P}}{\tilde{\bm{z}}_c^{\mathrm{H}}\tilde{\bm{z}}_c}},
\end{equation}
where $\bm{z} \in \mathbb{C}^k$ is the normalized signal that satisfies the power constraint, $(\cdot)^{\mathrm{H}}$ denotes the conjugate transpose, and $\mathrm{PN}(\cdot): \mathbb{C}^k \rightarrow \mathbb{C}^k$ represents the power normalization.}

Next, the normalized signal $\bm{z}$ is transmitted over the channel. Here, we define the bandwidth compression ratio (BCR) as $\rho = k/N$. A larger $\rho$ indicates more channel resources allocated per source sample, while a smaller $\rho$ reflects higher bandwidth efficiency.

Considering a complex additive white Gaussian noise (AWGN) channel, the received signal is modeled as
\begin{equation}
    \bm{\hat{z}}=\bm{z}+\bm{n},
\end{equation}
where $\bm{n} \sim \mathcal{CN}(0, \sigma^2 \mathbf{I})$ is the complex Gaussian noise. The channel SNR is defined as
\begin{equation}
   \mathrm{SNR} = 10\log_{10} \frac{\bar{P}}{\sigma^2} \text{dB}.
\end{equation}
At the receiver, the noisy complex-valued signal is first mapped back to the real-valued representation
\begin{equation}
\hat{\bm{z}}_r = \mathcal{C}^{-1}(\bm{\hat{z}}) \in \mathbb{R}^{2k}.
\end{equation}
A JSCC decoder $D(\cdot; \bm{\phi})$ then reconstructs the image $\hat{\bm{x}} \in \mathbb{R}^{3\times N_\text{H} \times N_\text{W}}$ from $\hat{\bm{z}}_r$, given by
\begin{equation}
\hat{\bm{x}} = D(\hat{\bm{z}}_r; \bm{\phi}),
\end{equation}
where $\bm{\phi}$ denotes the learnable parameters of the decoder. 


Conventionally, JSCC systems are trained by jointly optimizing the encoder and decoder parameters to minimize the expected reconstruction loss between the original image $\bm{x}$ and its reconstruction $\hat{\bm{x}}$, i.e.,
\begin{equation}
\min_{(\bm{\theta}, \bm{\phi})}  L\left(\bm{x}, \hat{\bm{x}} \right),
\end{equation}
where $L(\cdot,\cdot)$ denotes a reconstruction-quality loss function, such as distortion-based metrics (e.g., mean squared error, MSE), perceptual metrics (e.g., LPIPS), or a weighted combination of both. In the next section, we introduce a novel training-free JSCC framework that eliminates the need for end-to-end optimization of the encoder and decoder.

\subsection{SemanticStyleGAN}
SemanticStyleGAN \cite{SemanticStyleGAN} is a variant of GAN that enhances the standard GAN architecture by incorporating semantic segmentation supervision into the training process. This design enables different parts of the latent code to precisely control specific regions or attributes of the generated image, allowing targeted modifications without affecting other regions. Formally, let $G(\cdot; \mathcal{G})$ denote the generator parameterized by $\mathcal{G}$. The generation process of SemanticStyleGAN can be expressed as
\begin{equation}
\bm{x} = G(\bm{y}; \mathcal{G}),
\end{equation}
where $\bm{y} = [\bm{y}_0,\bm{y}_1,\ldots,\bm{y}_{N_S-1}]$ is the input latent code. Each semantic vector $\bm{y}_i \in \mathbb{R}^{N_L}$, of length $N_L$, corresponds to a specific semantic attribute or region in the image. \textcolor{black}{For example, in \cite{SemanticStyleGAN}, when SemanticStyleGAN is trained on the CelebAMask-HQ dataset \cite{CelebAMask-HQ}, it learns disentangled semantic vectors representing different facial attributes.  Specifically, $\bm{y}_0$ and $\bm{y}_1$ capture coarse information of the  entire image, while $\bm{y}_2$ to $\bm{y}_{24}$ control the details of the face, such as the shape of the eyes, eyebrows, and mouth.}

\begin{figure*}[!t]
    \centering
    \includegraphics[width=1\linewidth]{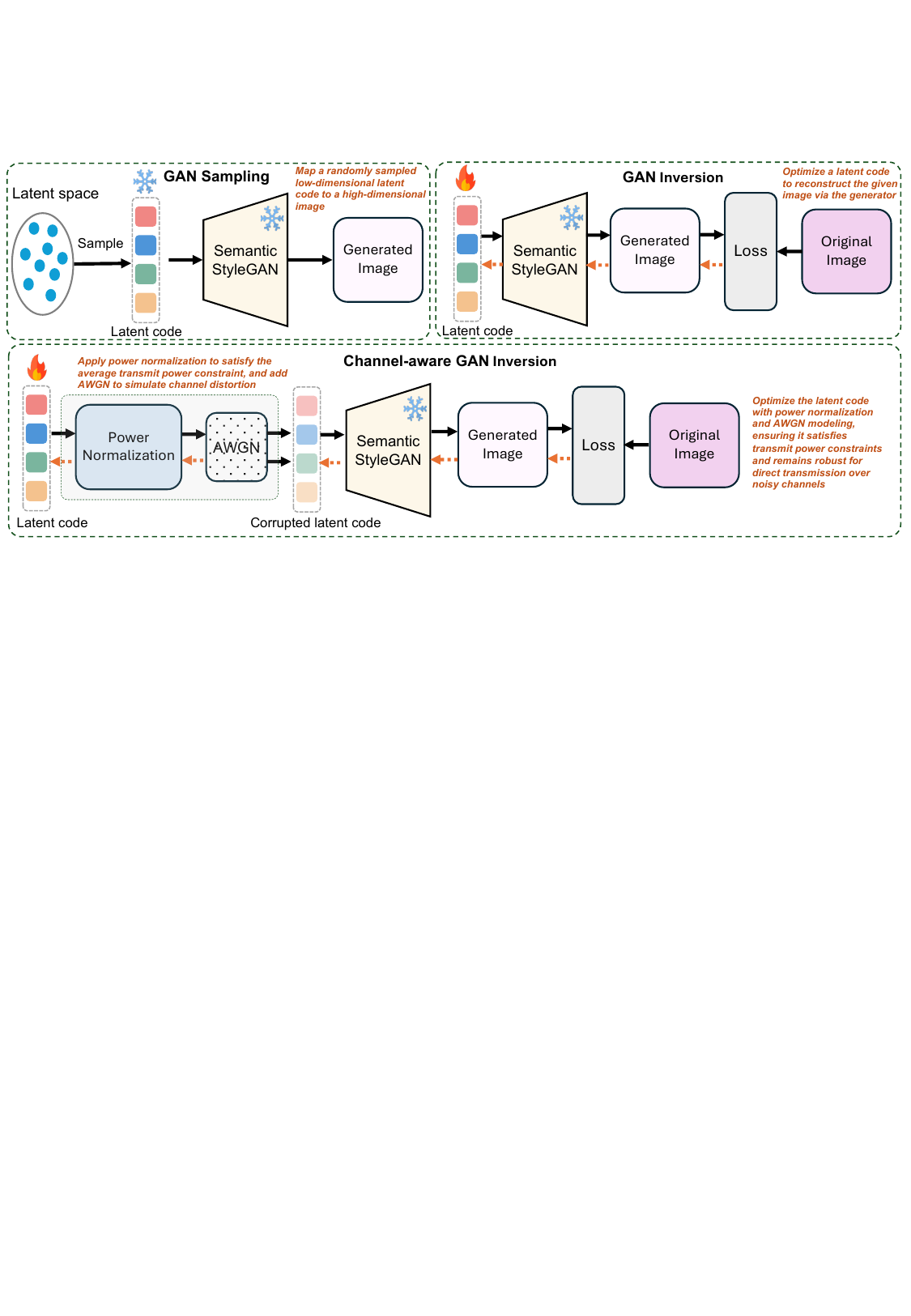}
    \caption{Illustration of the proposed channel-aware GAN inversion approach, where the channel characteristics are incorporated into the GAN inversion process to generate channel-adaptive semantic information. The derived semantic information then can be directly transmitted over noisy channels without explicit channel coding.}
    \label{fig:CAGI_overview}
\end{figure*}

\section{Channel-aware GAN Inversion based JSCC} 

In this section, we introduce the proposed CAGI-JSCC framework. We first present the GAN inversion method for semantic information extraction, followed by the proposed channel-aware GAN inversion process. Finally, we describe an improved optimization objective designed to further enhance the performance of CAGI-JSCC.

\subsection{GAN Inversion for Semantic Information Extraction}
Following our prior work \cite{tianxiao_GAN}, we apply GAN inversion to extract semantic information from the source image. As shown in \autoref{fig:CAGI_overview}, a well-trained GAN generates images by sampling a low-dimensional latent code from a prior distribution, whereas GAN inversion refers to the process of identifying a latent code that enables the generator to reconstruct an image visually similar to the original. In this context, the latent code obtained through inversion can be regarded as the semantic representation of the original image.

Mathematically, for a given image $\bm{x}$,  GAN inversion seeks a latent code $\bm{y}^*$ such that the generated image $G(\bm{y}^*;\mathcal{G})$ closely approximates $\bm{x}$. This objective can be formulated as
\begin{equation}
    \bm{y}^*  = \arg\min_{\bm{y}}||G(\bm{y};\mathcal{G}) - \bm{x}||^2_2,
    \label{eq::inversion_1}
\end{equation}
where the MSE loss is typically employed, and $G(\cdot;\mathcal{G})$ is the pre-trained generator with fixed parameters $\mathcal{G}$. To solve this optimization, one can initialize $\bm{y}$ by sampling from the prior distribution of the latent space and then iteratively update it using a gradient-based optimizer.

\subsection{Channel-aware GAN Inversion}
Once the latent code $\bm{y}^*$ is obtained, it cannot be directly transmitted after conversion into a complex-valued vector, since the resulting signal may violate the average power constraint. More critically, in the absence of protection or adaptation mechanisms such as channel coding, channel noise can severely distort the transmitted signal. A straightforward solution is to employ a learning-based channel encoder \cite{tianxiao_GAN, haowen_PGM}. However, such approaches require additional training, which prevents the direct use of pre-trained generative models in a training-free setting.

To overcome this limitation, we propose a channel-aware GAN inversion method that incorporates the effect of wireless channel characteristics into the inversion process, as shown in \autoref{fig:CAGI_overview}. We begin by sampling an initial latent code $\bm{y} \in \mathbb{R}^{2k}$ from the latent space. Instead of directly feeding $\bm{y}$ into the generator $G$, we first reshape it into a complex-valued vector $\tilde{\bm{z}} = \mathcal{C}(\bm{y})$ and \textcolor{black}{apply the power normalization operation in \eqref{eq::PN}} to ensure that the resulting signal satisfies the average power constraint. We assume that the transmitter has access to the current noise power level\footnote{More challenging scenarios are discussed in the simulation section, where the transmitter has no prior knowledge of the channel noise level.}. A noise vector $\bm{n}_A \sim \mathcal{CN}(0, \hat{\sigma}^2 \bm{I})$ is then sampled and added to the normalized signal to simulate the effect of the channel noise.

The overall channel-aware forward process can be given as
\begin{equation}
    A(\bm{y})= \mathcal{C}^{-1}\Bigg(\mathrm{PN}\Big(\mathcal{C}(\bm{y})\Big)+\bm{n}_A \Bigg),
\end{equation}
which maps the latent code to its noisy, power-constrained version in the real-valued latent space. Accordingly, the objective of channel-aware GAN inversion can be reformulated from \eqref{eq::inversion_1} as
\begin{equation}
    \bm{y}^* = \arg\min_{\bm{y}}||G\Big(A(\bm{y});\mathcal{G}\Big) - \bm{x}||^2_2.
    \label{eq::inversion_2}
\end{equation}
By solving this inversion problem with the channel-aware forward function, we obtain channel-adaptive semantic information that can be directly transmitted over noisy channels without additional training.

\begin{algorithm}[tbp]
\caption{Transmission pipeline of the Proposed CAGI-JSCC Approach}
\label{alg::CAGI_JSCC}
\textbf{Input:} Image $\bm{x}$, estimated noise power $\sigma^2$ \\
\textbf{Output:} Reconstructed image $\hat{\bm{x}}$ \\

\textbf{Transmitter:} \\
Perform GAN inversion without channel-aware processing by solving \eqref{eq::inversion_1} to obtain the initial latent code $\bm{y}$ of the input image $\bm{x}$. \\
Use $\bm{y}$ as the initialization to perform channel-aware GAN inversion by solving \eqref{eq::inversion_3} with the channel-aware forward function. Obtain the updated channel-aware latent code $\tilde{\bm{z}}$. \\
Convert $\tilde{\bm{z}}$ into a complex-valued and power-normalized transmit signal $\bm{z}=\mathrm{PN}\big(\mathcal{C}(\bm{y})\big)$ and transmit $\bm{z}$ over the wireless channel. \\

\textbf{Receiver:} \\
Receive the noisy signal $\bm{\hat{z}}$ from the channel. \\
Recover the real-valued latent code $\hat{\bm{z}}_r=\mathcal{C}^{-1}(\hat{\bm{z}})$. \\
Reconstruct the image $\hat{\bm{x}}$ by $G(\hat{\bm{z}}_r;\mathcal{G})$. \\
Return the reconstructed image $\hat{\bm{x}}$.
\end{algorithm}

\subsection{Improved Optimization Objective} 
To enhance reconstruction quality, we reformulate the inversion objective by integrating three complementary loss terms. First, we replace the MSE loss with the L1 loss. Compared with MSE, the L1 loss is more robust to extreme errors or outliers in pixel values \cite{WACV_loss,Mu2019CVPRW}. Since MSE squares the error, a few large deviations can dominate the optimization, often leading the model to average pixel values and produce overly smooth or blurred reconstructions. In contrast, the L1 loss increases linearly with the error, reducing the impact of outliers and enabling sharper structures and finer details to be preserved. In addition, we incorporate the LPIPS loss \cite{LPIPS} to capture perceptual similarity. LPIPS measures perceptual similarity between two images using deep features extracted from a pre-trained VGG network \cite{vgg}, and it is known to better align reconstruction quality with human visual perception.

We further introduce a task-specific loss $L_\text{task}$, which can be customized according to the downstream application. For example, in recognition tasks, it can be defined as classification error, while in segmentation tasks, it can measure the overlap between predicted and ground-truth masks. Accordingly, the overall optimization objective is formulated as
\begin{equation}
\begin{aligned}
    \bm{y}^* 
    &= \arg\min_{\bm{y}} \; 
       \mathcal{L}\!\Big( G\!\big(A(\bm{y});\mathcal{G}\big), \bm{x} \Big),
           \label{eq::inversion_3}
\end{aligned}
\end{equation}
where 
\begin{equation}
\begin{aligned}
      \label{eq::inversion_3_detail}
    \mathcal{L}\!\Big(G\!\big(A(\bm{y});\mathcal{G}\big), \bm{x}\Big) 
    &= \lambda_1 \,\Big\| G\!\big(A(\bm{y});\mathcal{G}\big) - \bm{x} \Big\|_1 \\
    &\quad + \lambda_2 \,\text{LPIPS}\!\Big(G\!\big(A(\bm{y});\mathcal{G}\big), \bm{x}\Big) \\
    &\quad + \lambda_3 \,L_\text{task}\!\Big(G\!\big(A(\bm{y});\mathcal{G}\big), \bm{x}\Big).
\end{aligned}
\end{equation}
Here, $\lambda_1$, $\lambda_2$, and $\lambda_3$ are positive hyperparameters that balance the contributions of the three loss terms.

\subsection{Transmission Pipeline}
As summarized in Algorithm \ref{alg::CAGI_JSCC}, the proposed CAGI-JSCC framework consists of three main steps. First, we perform the GAN inversion process without considering channel-aware processing by solving the optimization problem in \eqref{eq::inversion_1} to obtain the real-valued semantic latent code $\bm{y}$ of the input image $\bm{x}$. This initial latent code serves as the starting point for the subsequent channel-aware optimization. In the second step, we explicitly incorporate the effects of the transmit power constraint and channel noise by applying the forward function $A(\cdot)$ in the optimization process, using the improved objective in \eqref{eq::inversion_3}. Starting from the latent code obtained in the first step, we refine $\bm{y}$ to obtain a channel-aware latent representation that is robust to noisy channels.
Finally, the transmitter converts $\bm{y}$ into the normalized complex signal $\bm{z}=\mathrm{PN}\big(\mathcal{C}(\bm{y})\big)$ and transmits it over the wireless channel. After reception, the receiver maps $\bm{\hat{z}}$ back to the real-valued latent space via $\hat{\bm{z}}_r=\mathcal{C}^{-1}(\bm{\hat{z}}_c)$ and reconstructs the image $\hat{\bm{x}}$ by feeding $\hat{\bm{z}}_r$ into the generator $G(\cdot)$.

\begin{figure}
    \centering
    \includegraphics[width=1\linewidth]{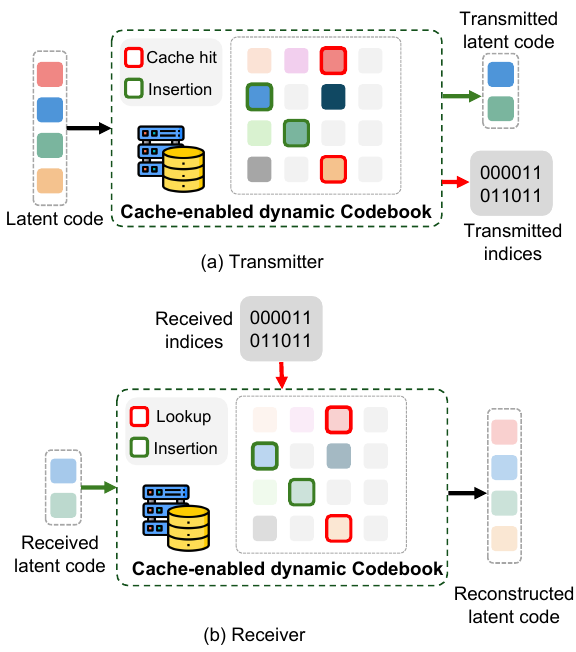}
    \caption{Illustration of the proposed CDC design, which deploys cache memories at both the transmitter and receiver to store previously transmitted semantic information and serve as a dynamic codebook.}
    \label{fig:CDC_overview}
\end{figure}
\section{Cache-enabled Dynamic Codebook}
\textcolor{black}{In this section, we will introduce the proposed CDC design to further enhance communication efficiency and then present how CDC can be integrated with the proposed CAGI-JSCC framework.}

\subsection{Cache-enabled Dynamic Codebook Design}
As shown in \autoref{fig:CDC_overview},   the transmitter is equipped with a cache memory organized by semantic attributes or regions, such as facial shape, eyes, ears. For each semantic vector $i \in \{0, \dots, N_S - 1\}$, the cache can store up to $N_C$ of them with length $N_L$, denoted by $\{\bm{c}_{i,0}, \dots, \bm{c}_{i,N_C-1}\}$, where $\bm{c}_{i,j} \in \mathbb{R}^{N_L}$. Accordingly, the total cache capacity is $N_S \times N_C$ semantic vectors, \textcolor{black}{each of size $N_L$}. The receiver maintains a cache of the same structure and size to enable synchronization.

At the transmitter, the cache stores previously transmitted latent codes and serves as a dynamic codebook. Upon obtaining $\bm{z}$ to be transmitted, the transmitter searches its local cache and checks whether similar semantic vectors exist.  If such semantic vectors exist, the transmitter replaces them with the corresponding indices of these vectors in the cache, thereby reducing the amount of information to be transmitted. For example, semantic vectors corresponding to attributes such as nose shape can be reused across different individuals if they exhibit similar semantics, allowing transmission of indices instead of full vectors.

Specifically, for semantic vector $\bm{z}_i$, the transmitter searches in its local cache memory for the closest cached semantic vector using cosine similarity, and obtains its index $j^*$ as follows
\begin{equation}
\begin{aligned}
    j^* =& \arg\max_{0\leq j < N_C } \cos(\bm{z}_i, \bm{c}_{i,j}) \\
    = &\arg\max_{0\leq j < N_C } \frac{\bm{z}_i\cdot \bm{c}_{i,j}}{|\bm{z}_i||\bm{c}_{i,j}|},
\end{aligned}
\end{equation}
where $\bm{c}_{i,j}$ denotes the $j$-th candidate in cache memory for the $i$-th semantic vector. If the cosine similarity $\cos(\bm{z}_i, \bm{c}_{i,j^*})$ is greater than a pre-defined threshold $\gamma_{i}$, we transmit the index $j^*$ instead of the original $\bm{z}_i$ to the receiver. \textcolor{black}{This operation is executed for all dimensions, resulting in a lower-dimensional semantic vector of length $n_s$ that excludes the cached semantic vectors and will be directly transmitted, denoted as $\tilde{\bm{z}}_{d} \in \mathbb{R}^{n_{s}\times N_L}$}, where
\begin{equation}
     n_s= N_S-\sum_{0 \leq i < N_S} \mathbb{I} \bigg(  \cos(\bm{z}_i, \bm{c}_{i,j^*})\geq \gamma_{i}\bigg),
\end{equation}
and $\mathbb{I}(\cdot)$ denotes the indicator function. 

For the semantic vectors that do not have a similar semantic vector in the cache, the transmitter will transmit them directly and store them in the cache memory for future use. For the indices of the cached semantic vectors, the transmitter transmits them using digital communication techniques. \textcolor{black}{More specifically, we can use a fixed-length binary code to represent the indices, where the number of bits required is $N_{\text{indices}}=\lceil \log_2 (N_C \times N_S) \rceil$ for each index, and we then apply channel coding and digital modulation to transmit the indices over the wireless channel reliably. With a channel code of rate $R_c$ and a modulation scheme carrying $M$ bits per symbol, each index requires $N_{\text{indices}}/{R_c M}$ channel symbols. In contrast, transmitting the original semantic vector via analog transmission requires $N_L/2$ complex channel symbols. Since $N_L$ is typically very large (e.g., $N_L = 512$ in SemanticStyleGAN) while $N_C$ and $N_S$ can be relatively small, the number of channel symbols required for index transmission is significantly smaller than that for transmitting the original semantic vector, thereby reducing the communication overhead.}

Upon receiving the directly transmitted semantic vectors $\tilde{\bm{z}}_{d}$ and indices, the receiver retrieves the corresponding semantic vectors in its local cache using the received indices, thus forming the complete latent code $\hat{\bm{z}}$. The receiver can then reconstruct the image $\hat{\bm{x}}$ by feeding $\hat{\bm{z}}$ into the generator $G(\cdot)$. Moreover, the receiver also stores the received semantic vectors in its local cache memory in the same manner as the transmitter. This ensures that both the transmitter and receiver maintain synchronized cache memories, allowing for efficient reuse of semantic vectors in future transmissions. It is important to note, however, that due to channel noise and other transmission imperfections, the actual values of the cached semantic vectors at the transmitter and receiver may differ.
\textcolor{black}{In addition, bit errors might occur during the transmission of indices, which leads to incorrect retrieval of cached semantic vectors at the receiver\footnote{\textcolor{black}{In this case, we adopt a similar fallback strategy used in \cite{DeepJSCC} for digital communications, where an invalid index is replaced by a randomly selected semantic vector.}}. We explicitly account for these potential discrepancies in our simulations, and the results shown in \autoref{Sec:sim} demonstrate that the proposed system still achieves significant improvements in overall reconstruction quality.}

\subsection{\textcolor{black}{SNR-aware Cache Update}}

\textcolor{black}{In practice, the cache memories at the transmitter and receiver do not have infinite storage capacity and it is necessary to design an effective cache update mechanism. One straightforward approach is to use the classic least recently used (LRU) policy, which evicts the least recently accessed items when the cache reaches its capacity. However, such a policy does not consider the impact of channel conditions during the transmission of semantic vectors. For example, some popular semantic vectors may be transmitted under poor channel conditions, resulting in degraded versions being stored in the cache. If these degraded semantic vectors are subsequently reused in future transmissions when the channel has improved, the quality of the reconstructed image may be significantly inferior compared to the system that transmits the original latent codes  when the channel condition gets better. Therefore, we propose an \textit{SNR-aware cache update mechanism} that takes into account the channel conditions during the transmission of semantic vectors. }

Specifically, we first associate each semantic vector with its corresponding transmit SNR at the time of transmission\footnote{The transmitter can obtain the current SNR through channel estimation based on pilot signals.}. When a new semantic vector $\bm{z}_i$ is to be stored in the cache, we check if there exists a cached vector $\bm{c}_{i,j^*}$ that is similar to $\bm{z}_i$. If such a vector exists, we compare the current SNR with the SNR associated with $\bm{c}_{i,j^*}$, denoted as $\text{SNR}_{i,j^*}$. If the current SNR is higher, which means that $\bm{z}_i$ is of higher quality than $\bm{c}_{i,j^*}$, we replace $\bm{c}_{i,j^*}$ with $\bm{z}_i$ in the cache, given by
\begin{equation}
\bm{c}_{i,j^*} \leftarrow
\begin{cases}
\bm{z}_i, & \text{if } \;  \text{SNR}_{\text{cur}} \geq \text{SNR}_{i,j^*}, \\[6pt]
\bm{c}_{i,j^*}, & \text{otherwise}.
\end{cases}
\end{equation}
where $\text{SNR}_{\text{cur}}$ is the current SNR when $\bm{z}_i$ is transmitted.
This ensures that the cache always contains the highest-quality version of each semantic vector. We refer to this process as \textit{SNR-aware cache upgrade}. Once such an upgrade is performed, we update the associated SNR value $\text{SNR}_{i,j^*} \leftarrow \text{SNR}_{\text{cur}}$ and transmit both the index $j^*$ and the fresh version of $\bm{z}_i$ to the receiver to ensure cache consistency.

Next, we design an SNR-aware eviction mechanism to handle the scenario when the cache reaches its maximum capacity. 
Similar to the classic LRU policy, we introduce an access timestamp $t(\bm{c}_{i,j})$ for each cached semantic vector $\bm{c}_{i,j}$, which is updated whenever a cache hit occurs. Specifically, the procedure can be described in the following steps. \textbf{Step 1}: for each cached semantic vector, we record its access timestamp $t(\bm{c}_{i,j})$ and its SNR tag $\text{SNR}_{i,j}$. \textbf{Step 2}: we then set a linear combination of the SNR tag and the access timestamp as the eviction priority score, given by
\begin{equation}
    \text{Priority}(\bm{c}_{i,j}) = \alpha \cdot \text{SNR}_{i,j} + (1-\alpha) \cdot t(\bm{c}_{i,j})
\end{equation}
where $\alpha \in [0,1]$ is a weight parameter that balances the importance of SNR and recency. \textbf{Step 3}: when the cache reaches its maximum capacity, we evict the cached semantic vector with the lowest priority score. This joint criterion ensures that the cache retains high-quality and frequently reused semantic vectors, thereby improving the quality of reconstructed content in future transmissions. We note that the values of $\alpha$ can be tuned based on the specific application requirements and the range of SNR values.

\subsection{Joint Optimization with CAGI-JSCC}
To better leverage the cache-enabled dynamic codebook, we further enhance the optimization objective in \eqref{eq::inversion_3} by considering the impact of the cache hit. This is because when some semantic vectors do not need to be transmitted, it is necessary to adjust the power allocated to the remaining semantic vectors to ensure that the average transmit power constraint is still satisfied. Moreover, cache hits introduce approximation errors, as the substituted cached vectors may not exactly match the original latent representations. To compensate for this, we introduce a distortion-aware process into the objective by giving additional attention to the non-replaced semantic vectors, allowing them to adapt in a way that mitigates these errors.

\begin{figure*}[t!]
    \centering
    \includegraphics[width=\linewidth]{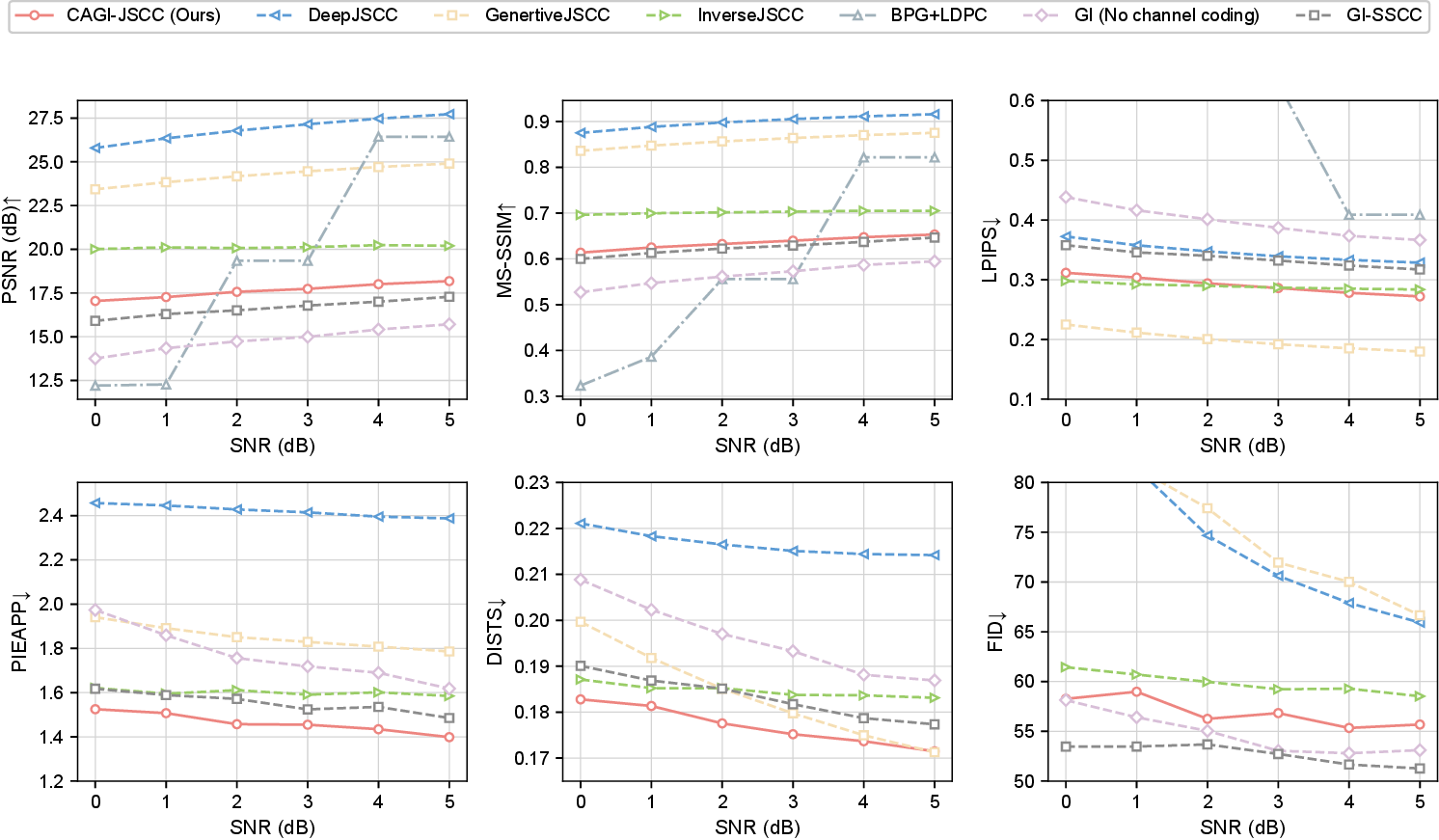}
    \caption{Performance comparison of the proposed CAGI-JSCC approach with various baseline methods in terms of PSNR, MS-SSIM, LPIPS, PIEAPP, DISTS, and FID versus SNR over the AWGN channel, where the BCR is set to $\rho=1/128$ and SNR varies from 0 dB to 5 dB.}
    \label{fig:performance_cagi}
\end{figure*}
Specifically, we first modify the forward function $A(\cdot)$ to incorporate the impact of the cache hit, which can be expressed as
\begin{equation}
    \tilde{A}(\bm{y})=C^{-1}\bigg(\mathcal{C}^{-1}\Big(\mathrm{PN}\big(\mathcal{C}(C(\bm{y})[1])\big)+\bm{n}\Big), C(\bm{y})[2]\bigg),
\end{equation} 
where $C(\cdot)$ denotes the aforementioned replacement operation that replaces the semantic vectors if a match is found in the transmitter-side cache. The output $C(\bm{y})[1]$ is the real-valued latent vector containing only the uncached semantic vectors, while $C(\bm{y})[2]$ stores the associated index list for reconstruction. After the reduced latent vector is transmitted through the complex channel, $\mathcal{C}^{-1}(\cdot)$ maps the received signal back to the real-valued latent space, and the inverse cache operation $C^{-1}(\cdot)$ restores the full-dimensional latent vector by combining the received components with the cache-retrieved semantic vectors.

We can then reformulate the optimization problem in \eqref{eq::inversion_3} as
\begin{equation}
    \bm{y}^*= \arg\min_{\bm{y}} \mathcal{L}\Big( G \big ({\tilde{A}(\bm{y})}; \mathcal{G}\big) , \bm{x}\Big).
    \label{eq::inversion_4}
\end{equation}
However, it is important to note that the caching operations $C(\cdot)$ and $C^{-1}(\cdot)$ are inherently non-differentiable, as they involve discrete indexing and memory access operations. This non-differentiability prevents the direct application of gradient-based optimization methods.

To address this issue, we use the straight-through operation to circumvent the non-differentiable caching operation during the gradient descent process. Specifically, we reformulate the optimization problem in \eqref{eq::inversion_4} as
\begin{equation}
    \bm{y}^*= \arg\min_{\bm{y}} \mathcal{L}\Big( G \big ({\bm{y}+\text{sg}[\tilde{A}(\bm{y})-\bm{y}]}; \mathcal{G}\big) , \bm{x}\Big),
\end{equation}
where $\text{sg}[\cdot]$ denotes the stop gradient function, indicating that gradients are not computed for the term within it during the back-propagation process. This technique, commonly used in neural network training involving non-differentiable operations \cite{Balle_Compression, VQ_VAE,DPQ}, enables the optimization problem to be solved using gradient descent methods.

\subsection{Two-stage Optimization}
The cache operation $C(\cdot)$ may eliminate a different set of semantic vectors in each iteration during the optimization, which introduces instability and hinders convergence. To address this issue, we propose a two-stage GAN inversion strategy. In the first stage, the optimization is performed without applying the cache operations $C(\cdot)$ and $C^{-1}(\cdot)$, allowing the latent space to be explored freely. At the beginning of the second stage, the cache operation $C(\cdot)$ is applied once to determine the cache hits, and the corresponding semantic vectors are then fixed throughout the subsequent optimization. The remaining semantic vectors are further refined to compensate for the approximation error introduced by the cached ones. This staged approach improves the optimization stability and leads to better reconstruction performance under cache-enabled transmission.

\section{Simulations}
\label{Sec:sim}
In this section, we present simulation results to evaluate the performance of the proposed CAGI-JSCC approach. Firstly, we outline the general simulation settings, including the dataset, evaluation metrics, and baseline methods. Then, we present the performance of the proposed CAGI-JSCC approach under different channel conditions. Moreover, we validate the effectiveness of the cache-enabled dynamic codebook design.

\subsection{General Setup}
In the simulations, we set the SNR of the AWGN channel to vary from 0 dB to 5 dB. The transmit power constraint is set to unit power, i.e., $\bar{P}=1$. The GAN model used in the simulations is the pre-trained SemanticStyleGAN \cite{SemanticStyleGAN} trained on the CelebAMask-HQ dataset \cite{CelebAMask-HQ}. The image resolution is set to $512 \times 512$, the latent code dimension is set to $N_L=512$, and the number of semantic vectors is set to $N_S=28$. For a fair comparison between the proposed CAGI-JSCC and the baseline methods, we select the first 24 semantic vectors as the transmitted semantic information, resulting in a BCR of $\rho = 1/128$ to align with the commonly used BCR for DeepJSCC architectures. To evaluate the performance of the proposed CDC design, we transmit all 28 semantic vectors. For the task-specific loss, we use the pre-trained FaceNet model\footnote{\url{https://github.com/timesler/facenet-pytorch}} to extract the identity features of the original and reconstructed images, and then compute the cosine similarity between these two features as the task-specific loss. The Adam optimizer is used to solve the optimization problems in \eqref{eq::inversion_1} and \eqref{eq::inversion_3}, with a learning rate of 0.01 and a maximum of 300 iterations. The hyperparameters in \eqref{eq::inversion_3} are set to $\lambda_1=0.1$, $\lambda_2=1$, and $\lambda_3=0.1$. \textcolor{black}{For transmitting the indices of the cached semantic vectors, we encode them using LDPC codes with a coding rate of 1/3, followed by BPSK modulation, and the total bandwidth cost for index transmission is also taken into account in the evaluation.} We report the performance of the proposed CAGI-JSCC approach in terms of the following metrics:
\begin{itemize}
    \item \textbf{Peak Signal-to-Noise Ratio (PSNR):} A distortion-based metric that quantifies the pixel-level difference between the original and reconstructed images. Higher PSNR values indicate better reconstruction quality.
    
    \item \textbf{Multiscale Structural Similarity Index Measure (MS-SSIM):} An extension of SSIM that evaluates structural similarity across multiple scales, providing a more robust assessment of perceptual quality.
    
    \item \textbf{LPIPS}\cite{LPIPS}: A perceptual metric that measures the similarity between two images based on deep network features. Lower values indicate higher perceptual similarity.
    \item \textbf{Perceptual Image-Error Assessment through Pairwise Preference (PIEAPP) \cite{PIEAPP}}: A learned perceptual error metric explicitly trained on large-scale human preference data. Unlike LPIPS, which compares feature distances, PIEAPP predicts human-perceived error directly. Lower PIEAPP values suggest that generated images are judged by humans to be closer to the reference.
    
    \item \textbf{Deep Image Structure and Texture
    Similarity  (DISTS)}\cite{DISTS}: A deep-learning-based metric that jointly evaluates structural and textural similarity, showing strong alignment with human quality assessments. Lower DISTS values indicate higher perceptual similarity.
    
    \item{\textbf{Fréchet Inception Distance (FID)}\cite{FID}}: A distribution-level metric that compares the statistics of real and generated images in a deep feature space. Lower FID values reflect closer alignment between the two distributions and thus better generative quality.
\end{itemize}
\begin{figure*}[!t]
    \centering
    \includegraphics[width=\linewidth]{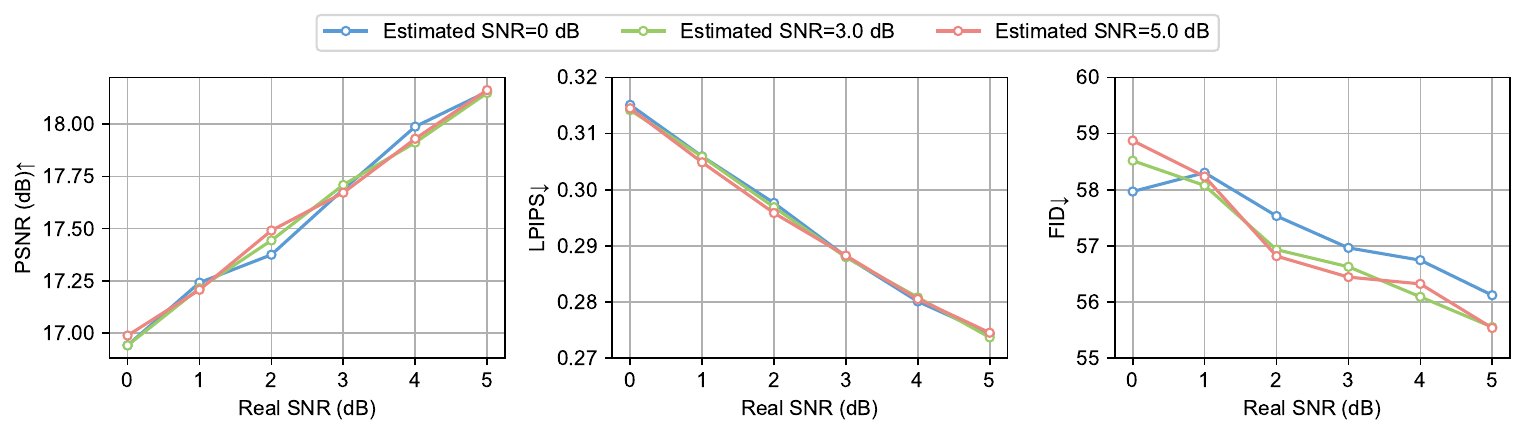}
    \caption{Performance of the proposed CAGI-JSCC approach with imperfect SNR knowledge at the transmitter, where the BCR is set to $\rho=1/128$ and real SNR varies from 0 dB to 5 dB.}
    \label{fig:unknown_snr}
\end{figure*}
    \begin{figure*}[t!]
    \centering
    \includegraphics[width=\linewidth]{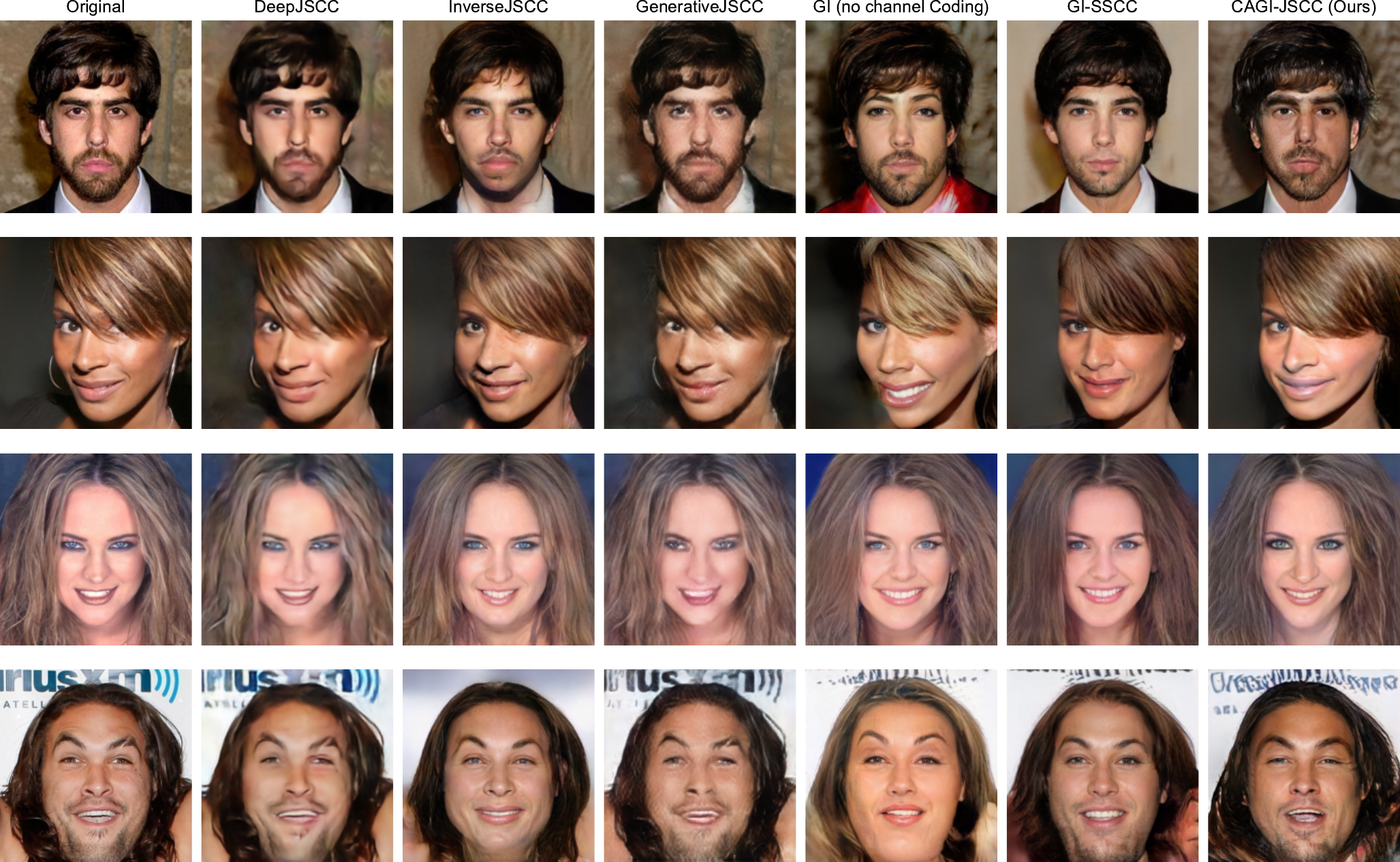}
    \caption{Visual comparison of different methods under SNR = 3 dB.}
    \label{fig:visual_CAGI}
\end{figure*}
Moreover, we choose the following baseline methods for comparison:
\begin{itemize}
    \item \textbf{DeepJSCC} \cite{DeepJSCC}: DeepJSCC is a learning-based JSCC approach that uses a deep neural network to learn the source and channel coding jointly.
    \item \textbf{InverseJSCC}\cite{GAN_JSCC}: InverseJSCC is a GAN-based JSCC approach that regards the transmission process of DeepJSCC as a degradation problem. It employs a pre-trained GAN at the receiver to iteratively solve the inverse problem of this degradation process, thereby reconstructing the original source data with high fidelity. For fair comparison, we use the same GAN model as in our proposed CAGI-JSCC approach.
    \item \textbf{GenerativeJSCC}\cite{GAN_JSCC}: Generative JSCC is also a GAN-based JSCC approach. Different from InverseJSCC, it jointly trains a DeepJSCC encoder and decoder with a frozen pre-trained GAN at the receiver. The output of the DeepJSCC decoder is fed into the GAN to reconstruct the image. We also use the same GAN model as in our proposed CAGI-JSCC approach.
    \item \textbf{GI-SSCC}: For the ablation study, we present the performance of using GAN inversion to extract the semantic information, followed by a learning-based JSCC encoder and decoder to transmit the extracted semantic information. The JSCC encoder and decoder are jointly optimized by minimizing the MSE between the transmitted and reconstructed latent codes.
    \item \textbf{GI (No channel coding)}: Following GI-SSCC, we eliminate the learning-based JSCC encoder and decoder, and directly transmit the extracted semantic information over the channel without any channel coding.
    \item \textbf{BPG + LDPC}: We also present the performance of a traditional separate source and channel coding scheme, where the image is first compressed using the BPG codec and then protected by the LDPC channel coding scheme. 
\end{itemize}

\subsection{Effectiveness of the proposed CAGI-JSCC}

\begin{figure*}[t!]
    \centering
    \includegraphics[width=\linewidth]{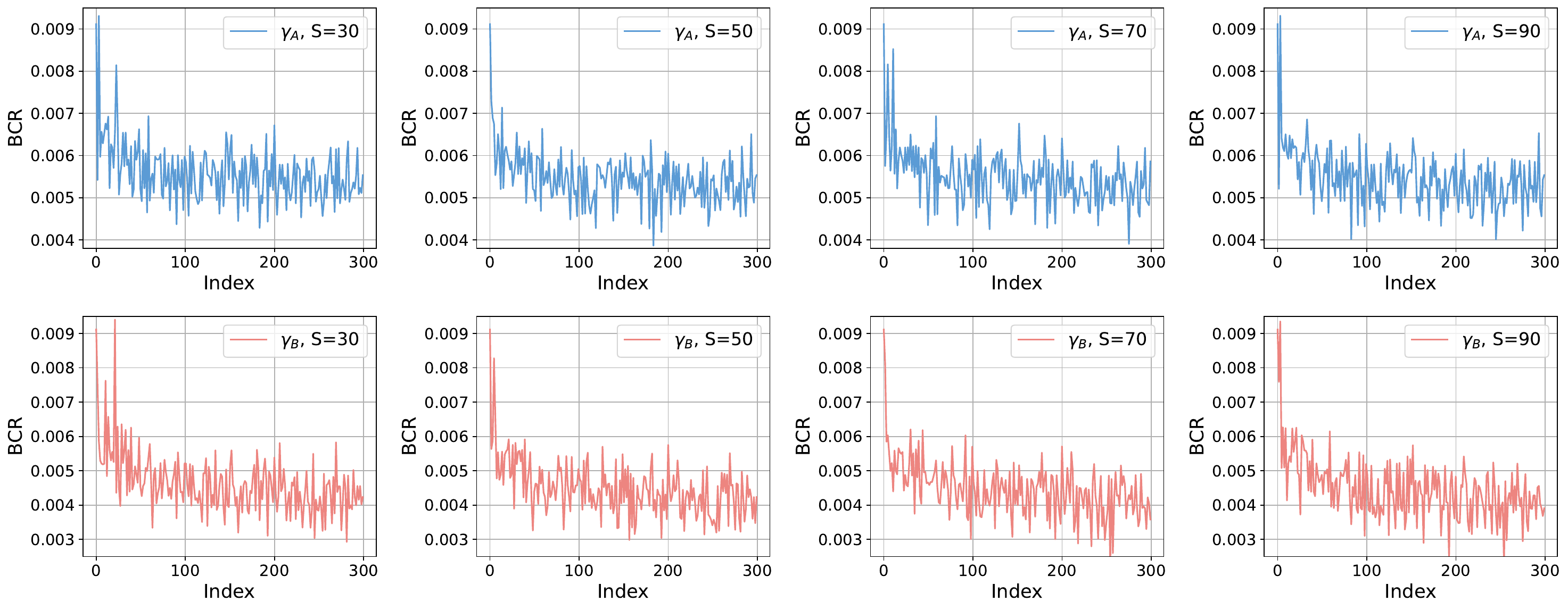}
    \caption{BCR versus the index of the transmitted images, where the SNR is randomly selected from 0 dB to 5 dB for each image transmission. The cache hit threshold is set to $\gamma_A$ and $\gamma_B$.}
    \label{fig:cdc_BCR}
    \end{figure*}
As shown in \autoref{fig:performance_cagi}, we first evaluate the performance of the proposed CAGI-JSCC approach without considering CDC, where SNR varies from 0 dB to 5 dB. From the results, we can observe that while DeepJSCC achieves the best performance in terms of PSNR and MS-SSIM, it falls short in metrics that assess the similarity from the perspective of feature space and human perception, including LPIPS, PIEAPP, DISTS, and FID. Compared to InverseJSCC and GenerativeJSCC, which also use the same GAN models, the proposed CAGI-JSCC approach demonstrates significant performance improvements in modern perceptual metrics, such as PIEAPP, DISTS, as well as distribution-based FID. Although GenerativeJSCC achieves the best LPIPS performance, it requires additional training of the DeepJSCC encoder and decoder, which limits its practical deployment. More importantly, such overemphasizing on LPIPS may lead to suboptimal performance in other perceptual metrics\cite{PD_tradeoff}, as also observed in the visual results of \cite{GAN_JSCC} and \autoref{fig:visual_CAGI}. 

We also compare the performance of the proposed CAGI-JSCC approach with GI-SSCC and GI (No channel coding) as ablation studies in \autoref{fig:performance_cagi}. From the results, we can observe that although GI-SSCC and GI (No channel coding) achieve better FID performance than the proposed CAGI-JSCC approach, they perform significantly worse in other perceptual metrics, such as LPIPS, PIEAPP, and DISTS. This is because FID measures the distribution similarity between real and generated images. When channel noise happens to share a similar distribution with the latent space prior, the FID score may remain relatively unaffected. However, due to the lack of channel-aware optimization, the generated images may exhibit noticeable differences with the original images in terms of perceptual quality, as reflected in other metrics. This observation highlights the importance of channel-aware optimization in enhancing the perceptual quality of reconstructed images in SemCom systems.

We also consider the scenario where the transmitter does not have accurate knowledge of the current noise power level. In this case, we set the estimated SNR of [0, 3, 5] dB and evaluate the performance of the proposed CAGI-JSCC approach with real SNR varying from 0 dB to 5 dB. As shown in \autoref{fig:unknown_snr}, it can be observed that the proposed CAGI-JSCC approach still achieves robust performance, even with imperfect SNR knowledge at the transmitter. Specifically, PSNR and LPIPS performance almost remain unchanged compared to the case with perfect SNR knowledge. As for FID, only a slight performance degradation is observed when SNR estimation errors exist. In particular, when the SNR estimation error reaches 5 dB, the FID increases by only about 1 compared with the case of perfect SNR estimation, demonstrating the robustness of the proposed method to imperfect channel knowledge.

As shown in \autoref{fig:visual_CAGI}, we present the visual comparison of different methods under an SNR of 3 dB. From this figure, we can clearly observe that the proposed CAGI-JSCC approach significantly outperforms DeepJSCC in terms of perceptual quality. DeepJSCC tends to produce overly smooth images that lack fine details and appear blurry. Moreover, the proposed CAGI-JSCC approach also demonstrates competitive, even better visual quality compared to InverseJSCC and GenerativeJSCC, which also use the same GAN model and need additional training. Compared to GI (No channel coding), the proposed CAGI-JSCC approach generates images with higher similarity to the original images, while GI (No channel coding) produces images with noticeable differences due to the lack of channel-aware optimization. Compared to GI-SSCC, the proposed CAGI-JSCC approach generates images with natural textures. This further validates the effectiveness of the proposed CAGI-JSCC approach in preserving the perceptual quality of the reconstructed images.

\subsection{Effectiveness of Cache-enabled Dynamic Codebook}

We then evaluate the performance of the proposed CDC design, where we consider two different cache hit thresholds, i.e., $\gamma_A$ and $\gamma_B$. The former is set to a relatively high value to ensure that only highly similar semantic vectors are replaced, while the latter is set to a lower value to allow more aggressive caching. The details of these two thresholds are provided in Appendix \ref{appendix:threshold}.
We also consider different cache memory sizes by varying $N_C$ and dynamic channel conditions, where SNR is randomly selected from 0 dB to 5 dB for each image transmission. The cache memory size is set to $N_C=30, 50, 70, 90$, respectively. We evaluate the average BCR over 300 image transmissions. 

\begin{table}[t]
\centering
\caption{Comparison of CDC-enabled (Cache) and non-CDC (CAGI-JSCC) methods under SNR = 5 dB. CDC reduces bandwidth while maintaining acceptable distortion and perceptual quality.}
\label{tab:cdc_comparison}
\begin{tabular}{lcccc}
\toprule 
\textbf{Method} & \textbf{Avg BCR} $\downarrow$ & \textbf{PSNR} $\uparrow$ &  \textbf{LPIPS} $\downarrow$ & \textbf{FID} $\downarrow$  \\
\midrule
non-CDC (SNR=5dB) & 1/128 & \textbf{18.18}  & \textbf{0.2721} & 55.68 \\
\midrule
CDC ($\gamma_A$, $N_C=30$)  & 1/183 & 15.71 & 0.3341 &  56.66\\
CDC ($\gamma_A$, $N_C=50$)  & 1/184 & 16.18 & 0.3226 &  56.43\\
CDC ($\gamma_A$, $N_C=70$) & 1/182 & 16.32 &  0.3203 &  \textbf{54.87}\\
CDC ($\gamma_A,$ $N_C=90$) & 1/183 & 16.24 &  0.3206 & 56.33 \\
\midrule
CDC ($\gamma_B$, $N_C=30$)& 1/217 & 16.03 &  0.3303 & 56.55 \\
CDC ($\gamma_B$, $N_C=50$)  &  1/223& 16.18 &  0.3357 &  59.69\\
CDC ($\gamma_B$, $N_C=70$) & 1/224& 16.32 &  0.3446 & 62.31 \\
CDC ($\gamma_B$, $N_C=90$) & 1/224 & 16.24 &  0.3438 & 60.24 \\
\bottomrule
\end{tabular}
\end{table}
As shown in \autoref{fig:cdc_BCR}, for the initial transmission with empty cache memories, the BCR for the first image by the proposed CDC is approximately $1/128$. As the cache memories at the transmitter and receiver are populated with semantic vectors, the BCR quickly decreases. Specifically, with thresholds set to $\gamma_A$ in our proposed system, the smallest BCR achieves $1/256$. When thresholds shift to $\gamma_B$, the BCR declines even more swiftly, reaching a value of $1/1024$ with only a few semantic vectors transmitted. This can be attributed to the proposed CDC, by which transmitted semantic vectors are cached and utilized to replace similar ones in the semantic vectors of subsequent transmissions, thus avoiding redundant transmissions and enhancing communication efficiency. Notably, we observe fluctuations in the BCR values, which arise from certain images exhibiting more similarities in their semantic vectors with previously transmitted images. These results demonstrate the evolving characteristics and the effectiveness of the proposed system.

In \autoref{tab:cdc_comparison}, we present the numerical results comparing the performance of the proposed CDC design with different cache hit thresholds and cache memory sizes. To provide a fair comparison, we also include the performance of the CAGI-JSCC approach without caching under SNR = 5 dB, where BCR is set to $1/128$. For the proposed CDC, we still consider the dynamic channel conditions, where SNR is randomly selected from 0 dB to 5 dB for each image transmission. As shown in the table, the non-CDC baseline achieves a PSNR of 18.18 dB and an FID of 55.68 with a BCR of $1/128$, while the proposed CDC design with different configurations substantially reduces the BCR to as low as $1/224$, indicating a significant improvement in transmission efficiency. Meanwhile, the degradation in image quality remains minimal across all metrics. Notably, under configuration $(\gamma_A, N_C=70)$, the system achieves the best FID 54.87, even outperforming the non-caching baseline, while maintaining a much lower BCR of $1/182$.

Besides, it can be seen that a more aggressive cache threshold $\gamma_B$ leads to higher communication efficiency by allowing more cached semantic vectors to be reused. However, increasing the cache size $N_C$ can only bring slight improvements. This is because the reuse frequency of semantic vectors follows a long-tail distribution\cite{DBLP:conf/cvpr/ZhongDWHPTH19,DBLP:conf/cvpr/CaoZHGL20}, where a small number of highly frequent semantic vectors are repeatedly encountered and quickly cached, while the remaining ones occur rarely and contribute little to further reuse. Therefore, a moderate cache size is sufficient for the proposed CDC scheme. These results can demonstrate the effectiveness and robustness of the proposed CDC mechanism in achieving a good trade-off between transmission efficiency and reconstruction quality under varying caching configurations.

\begin{figure}[t!]
    \centering
    \includegraphics[width=\linewidth]{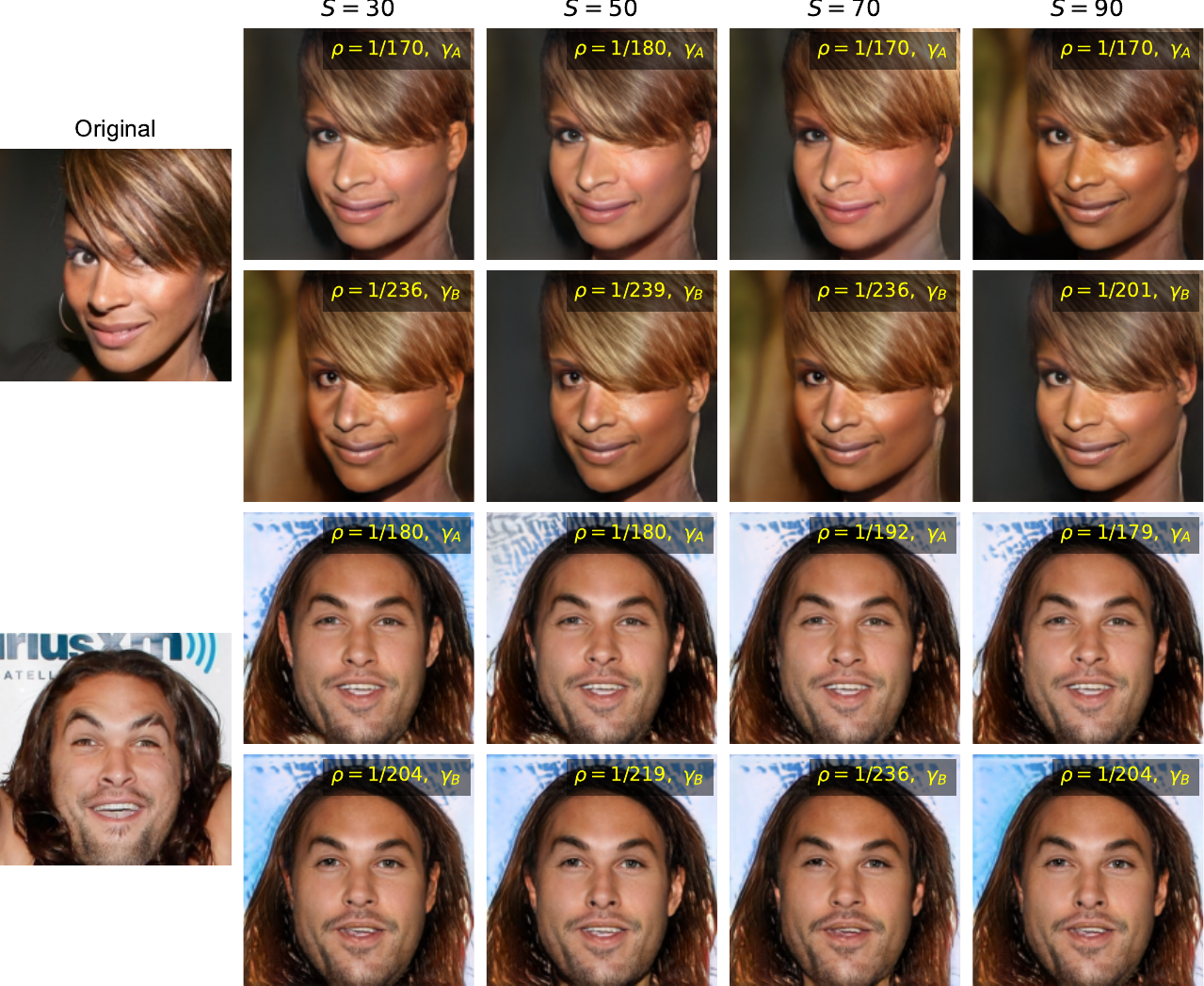}
    \caption{Visual comparison of the proposed CDC design with different cache hit thresholds and cache memory sizes.}
    \label{fig:visual_cdc}
\end{figure}

As shown in \autoref{fig:visual_cdc}, we present the visual results of the proposed CDC design with different cache hit thresholds and cache memory sizes. From this figure, it is evident that the proposed CDC design consistently produces high-quality reconstructed images across various configurations. Despite the reduction in BCR, the facial details and textures are well-preserved and only the background with less semantic importance exhibits some degradation. These results further validate the effectiveness of the proposed CDC design in reducing communication overhead while preserving the perceptual quality of the reconstructed images.

\section{Conclusion}
In this paper, we have investigated an evolving SemCom system designed to continually enhance its transmission efficiency by leveraging accumulated knowledge from previous transmissions stored in cache memories at both the transmitter and receiver. Specifically, we proposed a novel CAGI approach for JSCC, which takes into account channel noise and power normalization during the GAN inversion process. This method effectively maps input images into a disentangled channel-correlated latent space, enabling direct transmission over noisy channels without the need for additional channel encoding, thereby significantly improving system robustness under diverse channel conditions. Moreover, we introduced a CDC mechanism that deploys cache memories at both the transmitter and receiver to store previously transmitted semantic vectors. These cached vectors are then utilized to replace similar semantic vectors in subsequent transmissions, thereby reducing communication overhead. We also proposed an SNR-aware cache update to maintain cache consistency and quality under varying channel conditions. Simulation results demonstrate the evolving capability and superiority of our proposed system in terms of the communication efficiency and the perceptual quality of reconstructed images compared to both traditional methods and existing SemCom approaches.
\appendix
\section{Details of similarity threshold} \label{appendix:threshold}
We present the values of $\gamma_i$ for different semantic vectors in \autoref{tab:2} and \autoref{tab:3}, which are used in the proposed semantic caching mechanism. The value of $\gamma_i$ is determined by the extent of impact a semantic vector has on the perceptual quality of images. Specifically, we set a higher threshold for semantic vectors deemed more crucial for image reconstruction, and a lower threshold for those considered less important.

\begin{table*}
    \centering
    \caption{The value of $\gamma_i$ for different semantic vectors in $\gamma_A$.}
    \label{tab:2}
    
    \begin{tabularx}{\linewidth}{p{1.3cm}<{\centering} p{2cm}<{\centering} p{1.3cm}<{\centering} || p{1.3cm}<{\centering} p{2cm}<{\centering} p{1.3cm}<{\centering} ||p{1.3cm}<{\centering} p{2cm}<{\centering} p{1.3cm}<{\centering} }
    \toprule
    \textbf{No.} & \textbf{Facial areas} & \textbf{Threshold} &  \textbf{No.} & \textbf{Facial areas} & \textbf{Threshold} &  \textbf{No.} & \textbf{Facial areas} & \textbf{Threshold}\\
    \midrule
    $\gamma_0$ & Coarse Profile 1 & 0.90 &$\gamma_{10}$   &   Mouth Shape& 0.95 & $\gamma_{20}$   & Cloth Shape & 0.80  \\

    $\gamma_1$ & Coarse Profile 2 & 0.95 &$\gamma_{11}$ & Mouth Texture  & 0.95 & $\gamma_{21}$   &  Cloth Texture & 0.80  \\

    $\gamma_2$ & Background 1     & 0.80 &$\gamma_{12}$ &  Nose Shape  & 0.90 & $\gamma_{22}$   & Glass & 0.80  \\

    $\gamma_3$ & Background 2     & 0.80 &$\gamma_{13}$ & Nose Texture  & 0.90 & $\gamma_{23}$   & Unknown 1 & 0.10  \\

    $\gamma_4$ & Face Shape       & 0.95 &$\gamma_{14}$ &  Ear Shape & 0.85 & $\gamma_{24}$   & Hat & 0.50  \\

    $\gamma_5$ & Face Texture     & 0.95 &$\gamma_{15}$ &  Ear Texture & 0.85 & $\gamma_{25}$   & Unknown 2 & 0.10  \\

    $\gamma_6$ & Eye Shape        & 0.95 &$\gamma_{16}$ & Hair Shape  & 0.90 & $\gamma_{26}$   & Earring & 0.80  \\

    $\gamma_7$ & Eye Texture           & 0.95 &$\gamma_{17}$ & Hair Texture  & 0.90 & $\gamma_{27}$   & Unknown 3 & 0.10  \\

    $\gamma_8$ & Eyebrow Shape         & 0.95 &$\gamma_{18}$ &Neck Shape& 0.90 &  -  & - &  - \\

    $\gamma_9$ & Eyebrow Texture       & 0.95 &$\gamma_{19}$ &   Neck Texture  & 0.90 &  -  & -  & - \\
    \bottomrule
    \end{tabularx}
\end{table*}
\begin{table*}
    \centering
    \caption{The value of $\gamma_i$ for different semantic vectors in $\gamma_B$.}
    \label{tab:3}
    
    \begin{tabularx}{\linewidth}{p{1.3cm}<{\centering} p{2cm}<{\centering} p{1.3cm}<{\centering} || p{1.3cm}<{\centering} p{2cm}<{\centering} p{1.3cm}<{\centering} ||p{1.3cm}<{\centering} p{2cm}<{\centering} p{1.3cm}<{\centering} }
    \toprule
    \textbf{No.} & \textbf{Facial areas} & \textbf{Threshold} &  \textbf{No.} & \textbf{Facial areas} & \textbf{Threshold} &  \textbf{No.} & \textbf{Facial areas} & \textbf{Threshold}\\
    \midrule
    $\gamma_0$ & Coarse Profile 1 & 0.85 &$\gamma_{10}$   &   Mouth Shape& 0.92 & $\gamma_{20}$   & Cloth Shape & 0.75  \\

    $\gamma_1$ & Coarse Profile 2 & 0.85 &$\gamma_{11}$ & Mouth Texture  & 0.92 & $\gamma_{21}$   &  Cloth Texture & 0.75  \\

    $\gamma_2$ & Background 1     & 0.80 &$\gamma_{12}$ &  Nose Shape  & 0.85 & $\gamma_{22}$   & Glass & 0.75  \\

    $\gamma_3$ & Background 2     & 0.80 &$\gamma_{13}$ & Nose Texture  & 0.80 & $\gamma_{23}$   & Unknown 1 & 0.10  \\

    $\gamma_4$ & Face Shape       & 0.92 &$\gamma_{14}$ &  Ear Shape & 0.80 & $\gamma_{24}$   & Hat & 0.50  \\

    $\gamma_5$ & Face Texture     & 0.92 &$\gamma_{15}$ &  Ear Texture & 0.80 & $\gamma_{25}$   & Unknown 2 & 0.10  \\

    $\gamma_6$ & Eye Shape        & 0.92 &$\gamma_{16}$ & Hair Shape  & 0.85 & $\gamma_{26}$   & Earring & 0.75  \\

    $\gamma_7$ & Eye Texture           & 0.92 &$\gamma_{17}$ & Hair Texture  & 0.85 & $\gamma_{27}$   & Unknown 3 & 0.10  \\

    $\gamma_8$ & Eyebrow Shape         & 0.92 &$\gamma_{18}$ &Neck Shape& 0.85 &  -  & - &  - \\

    $\gamma_9$ & Eyebrow Texture       & 0.92 &$\gamma_{19}$ &   Neck Texture  & 0.85 &  -  & -  & - \\
    \bottomrule
    \end{tabularx}
\end{table*}
\bibliographystyle{IEEEtran}
\bibliography{IEEEabrv, references}

\end{document}